\documentclass[fleqn,10pt]{wlscirep}
\usepackage{cases}
\title{Quantum Scissor from Exact Generalized Photon Number Statistics}

\author[1]{Abdul Q Batin}
\author[2]{Suranjana Ghosh}
\author[3]{Prasanta K. Panigrahi}

\author[4,*]{Utpal Roy}

\affil[1,4]{Department of Physics, Indian Institute of Technology Patna, Bihta, Patna-800013, India}
\affil[2, 3]{Indian Institute of Science Education and Research Kolkata, Mohanpur-741246, India}
\affil[3] {Centre for Quantum Science and Technology (CQST), Siksha ‘O’ Anusandhan Deemed to be University, Khandagiri Square, Bhubaneswar-751030, Odisha, India}
\affil[*]{uroy@iitp.ac.in}

\keywords{Generalized coherent state, Generalized photon-added coherent state, Fock state}

\begin{abstract}
We report the close form expressions of the photon number statistics for a generalized coherent state and a generalized photon-added coherent state, which are shown to be crucial for proposing a variety of quantum scissor operations. The analytically obtained distributions are also capable of predicting the precise laser intensity windows for realizing a variety of quantum scissors. Truncating a photon added state overcomes the selection rule of obtaining the lower order Fock states. Photon addition also enables us to obtain a higher order Fock state in a lower order superposition. The importance of circular geometry is also demonstrated for engineering such quantum scissors.
\end{abstract}

\begin{document}

\flushbottom
\maketitle



\section*{Introduction}
In 1926, Erwin Schr{\"o}dinger first introduced the concept of coherent state (CS) through a non-spreading wavepacket of harmonic oscillator. The state is the most classical state in quantum domain, having Poisson photon number statistics, where the average photon number is directly correlated to the absolute square of the CS parameter, $\alpha$.
Though CS behaves classically, the quantum supremacy can be accomplished by quantum superposition, entanglement, squeezing, photon anti-bunching, negativity of Wigner distribution, \emph{etc} \cite{sanders2012review,sun1992generation,
yurke1986generating,janszky1990squeezing,orszag1992superposition,
buvzek1991origin,hillery1987amplitude,agarwal1992new,
solomon1994characteristic,wigner1997quantum}. Such quantum featured states are paid huge attention for their breakthrough contributions in diverse fields and possible technological implications \cite{nielsen2002quantum,jeong2001quantum,ralph2003quantum,bergmann2016quantum,
lee2005generation,lee2005quantum,chuang2015quantum,chuang2021generation}. In addition, quantum entanglement and teleportation \cite{deb2002tripartite,deb2003entangling,deb2008entanglement,
horoshko2016entanglement,horoshko2019quantum}, nonlinear dynamics of coherent field
\cite{suelzer2016effects,fischer2004experimental}, quantum sensing and metrology \cite{ghosh2006mesoscopic,ghosh2009sub,roy2009sub,ghosh2014enhanced,
ghosh2019sub,akhtar2021sub,agarwal2022quantifying,
bera2020matter,bera2022quantum,seveso2020quantum,
garbe2020critical,bonalda2019quantum} are the subjects of some of the recent emphasizes.

Often, the nonlocal superposition of two CSs, called Schr{\"o}dinger cat state (SCS), sets the building block of the quantum information protocols \cite{albert2016holonomic,asboth2004coherent,
jeong2003quantum,lund2008fault}. Several experimental schemes exist to create higher order mesoscopic superposition state in experiments \cite{Miranowicz_1990,praxmeyer2007time,praxmeyer2016direct,vogel2000nonclassical,
mikheev2019efficient,tyagi2021photon,choudhury2011proposal,rivera2022strong,
thekkadath2020engineering,molnar2018quantum}. Unlike a CS, a photon added coherent state (PACS), introduced by Agarwal and Tara, is non-Gaussian and manifests nonclassical nature \cite{agarwal1991nonclassical}. Their experimental generations have been reported \cite{zavatta2004quantum,yan2007optical,ramos2014engineering,
li2018generation,shringarpure2019generating,francis2020photon} and nonclassical behaviours are thoroughly investigated \cite{zavatta2005single,hong2010generalized,ren2019entanglement,tyagi2021photon}. PACS is also shown to be a favourable candidate for technological aspects like quantum sensing \cite{braun2014precision}, quantum key distribution \cite{barnett2018statistics} \emph{etc}.

In this paper, we analytically derive a close form expressions of the photon number distribution (PND) functions for a generalized-CS (GCS) and a generalized-PACS (GPACS). Apart from the extensive merit of this PND in quantum optics, here we use the PND for truncating the quantum states, termed as quantum scissors, as proposed by Pegg \textit{et al.} \cite{Pegg}. Quantum Scissors have found enormous importance in the literature and various experimental schemes for quantum state truncation are also in place \cite{Pegg1,PhysRevA.64.063818,koniorczyk2000general,leonskiquantum,
lee2010quantum,leonski2011quantum,abo2022hybrid,wang2022statistical}. In this case, the said truncation can be achieved based on the knowledge of the reported photon distributions. It is well-known to have the photon distribution of a CS as Poissonian. However, a superpositions of CSs are non-Gaussian and quantum, having super- or sub-Poisson-distribution. Here, we show that, the envelop of the obtained photon distribution takes the form of a Poissonian and this is valid for any order ($N$) of superposition, making it a universal feature of the state to decide upon the parameter domains for various quantum scissors. The scissor operation is demonstrated for obtaining vacuum state, other Fock states, equal superposition of Fock states, and mesoscopic quantum superposition states. Such crucial prediction is not possible without the exact knowledge of the parameter restriction, imposed by the reported photon statistics. Regarding truncating to a Fock state from a GCS, the intriguing fact being, not to get a Fock state below a certain $|n \rangle$, denoted by $|n_c \rangle$, whatsoever be the value of $N$. In addition, experimental generation of the Fock state $|n (>n_c)\rangle$ will only be possible for a specific range of $\alpha$ ($=\triangle \alpha$).

In addition to the quantum scissor operation for a GCS, we also demonstrate the scissor operations for a GPACS. Interestingly, GPACS allows generation of Fock states with $|n (< n_c) \rangle$, which is not possible for GCS. Moreover, a $r$-photon added GPACS is capable of generating a particular Fock state from a relatively lower order superposition. However, this reduction is allowed only above a certain order of superposition, which is $N=6$, the hexagonal state, also known as benzene-like state \cite{roy2009sub}. We also demonstrate the production of equal Fock states superposition and identify the precise allowed parameter domains for the above quantum states (but not limited to). Moreover, we prove that the primary requirement of these comprehensive outcomes relies on the merit of symmetrical circular arrangement, a slight deviation of which (as illustrated by an elliptic state) will significantly aggravate the scissor operations.

The paper is organized as follows. In the next section, we present the analytical derivation of the modulated photon distribution for a GCS, which is extended for a GPACS in Sec.III. In Sec.IV, a couple of examples of the quantum scissor operations are demonstrated. Precise parameter domains are identified for each generation in Sec.V. A deviation from the circular geometry is adopted in elliptic form in Sec.VI. We conclude in Sec.VII with the summary of outcomes and possible future directions.

\section*{Analytical Derivation of the PND for a GCS}

A single coherent state is known to exhibit Poissonian photon distribution, where consecutive energy levels contribute, which is not the case for GCS. We will start by writing the general form of a GCS by \cite{ghosh2015mesoscopic}
\begin{eqnarray}
|\psi\big\rangle=\mathcal{N}\frac{1}{\sqrt{N}} \sum_{j=1}^N|\alpha e^{\frac{i2\pi j}{N}}\big\rangle,
\label{wavefunction}
\end{eqnarray}
where the normalization constant, $\mathcal{N}$, is expressed as
\begin{eqnarray}
\mathcal{N} =\sqrt{N} \left[\sum_{j_{1},j_{2}=1}^N exp\left(\alpha_{j_1} \alpha^{\ast}_{j_2}-|\alpha|^{2}\right)\right]^{-\frac{1}{2}},
\label{Normalization}
\end{eqnarray}
where $\alpha_{j}=|\alpha|e^{\frac{i2\pi j}{N}}$. This constant $\mathcal{N}$ is for a GCS, which goes to unity for a CS and for $j_1=j_2$. This also provides the analytical expressions of the normalization constants for the quantum superposition states of lower orders \cite{gerry1997quantum,dong2000superposition}. We intend to derive a close form expression of the photon number distribution (PND) for the above state and the probability of the $n$-th state ($P_n$) becomes
\begin{eqnarray}
P_n && =  |\big\langle n|\psi\rangle |^2 \nonumber\\
&& =\frac{|\mathcal{N}|^{2}}{N}  e^{-|\alpha|^2}\frac{|\alpha| ^{2n}}{n!}\sum_{j_{1},j_{2}=1}^{N}e^{\frac{i2\pi n(j_1-j_2)}{N}}\nonumber\\
&& =\frac{|\mathcal{N}|^{2}}{N} e^{-|\alpha|^2}\frac{|\alpha| ^{2n}}{n!}\left[N+\sum_{j_1, j_2(\neq j_1)=1}^{N}e^{\frac{i2\pi n(j_1-j_2)}{N}}\right].\nonumber
\end{eqnarray}
Here, the diagonal terms are taken out from the summation and then, the term inside summation is converted into trigonometric function:
\begin{eqnarray}
P_n \! =\!\frac{|\mathcal{N}|^{2}}{N} e^{-|\alpha|^2}\frac{|\alpha| ^{2n}}{n!}\!\!\left[N \!+\!2\!\!\!\!\!\!\sum_{j_1,j_2(> j_1)=1,2}^{N-1,N} \!\!\!\!\! \cos{\frac{2\pi n(j_1-j_2)}{N}}\right].\nonumber
\end{eqnarray}
The above equation can be simplified for single summation index, $m=|j_1-j_2|=1,\;2,\;3,\;.....,\;(N-1)$ as
\begin{eqnarray}
P_{n} =\frac{|\mathcal{N}|^{2}}{N} e^{-|\alpha|^2}\frac{|\alpha|^{2n}}{n!}
\left[N+2 \sum_{m=1}^{N-1} (N-m)\cos{\frac{2\pi nm}{N}}\right].
\label{PND one}
\end{eqnarray}
For arriving at a definite expression of $P_n$, the term inside the square bracket needs to be simplified further. It involves the trigonometric summations in the forms, $\sum_{m=1}^{N-1} \cos{m x}$ and $\sum_{m=1}^{N-1} m \cos{m x}$, where $x = 2\pi n/N$. These can be converted to a term without summation by using the identities from the book by I. S. Gradshteyn and I. M. Ryzhik (sec. $1.342$) \cite{zwillinger2007table}. After a straight-forward simplification, the square bracketed term of Eq.(\ref{PND one}) takes the form, $\left[\sin(n\pi)/ \sin\Big(\frac{n\pi}{N}\Big)\right]^{2}$, which enables us to write PND in a much simpler form:
\begin{eqnarray}
P_{n} =\frac{|\mathcal{N}|^{2}}{N} e^{-|\alpha|^2}\frac{|\alpha|^{2n}}{n!}
\left[\frac{\sin(n\pi)}{\sin\Big(\frac{n\pi}{N}\Big)} \right]^{2}.
\label{PND condition}
\end{eqnarray}

It is worth noticing that, the numerator vanishes ($\sin(n\pi)=0$) for all $n$. This may give an impression that, PND vanishes for all energy levels, which is neither physically, nor mathematically justified. The reason being the value of the denominator, \emph{i.e.}, $\sin(\frac{n\pi}{N})$, which becomes zero for the levels, $n=S\times N$. Thus, we obtain two parameter domains:\\
   \textbf{I.} For $n\neq S\times N$, PND will simply vanish for all $n$ , where $\sin(n\pi)=0$, but  $\sin (\frac{n \pi}{N}) \neq 0$.\\
   \textbf{II} For $n = S \times N$, both the numerator and the denominator become zero, making the PND into a (zero/zero) form or indeterminate form. \\
   In such situation, the L’H$\hat{o}$pital’s rule is used to find out the value of the function at those point, given by $n=S \times N$. According to the L’H$\hat{o}$pital’s rule, if \textit{f(y)} and \textit{g(y)} are both differentiable at $y=a$, and also the function, $\emph{f(y)}/\emph{g(y)}$ at $y=a$ takes a ($zero/zero$) form, then the value of the function at $y=a$ can be evaluated by taking the derivatives, such that
\begin{eqnarray}
\lim_{y \to a} \frac{f(y)}{g(y)} = \lim_{y \to a} \frac{f^\prime(y)}{g^\prime(y)}=\frac{f^\prime(a)}{g^\prime(a)}.
\end{eqnarray}

This is applied in Eq. (\ref{PND condition}) for $y=n$ and $a=SN$. The value of the function at $n=SN$  becomes
\begin{eqnarray}
\lim_{n \to SN} \frac{\sin{n\pi}}{\sin{\frac{n\pi}{N}}} && = \lim_{n \to SN} \left[\frac{\pi \cos{(n\pi)}}{(\frac{\pi}{N})\cos (\frac{n\pi}{N})}\right]\nonumber\\
&& =N.
\end{eqnarray}
This facilitates us to write the final form of the PND in a simple algebraic form involving $\alpha$ without a trigonometric functional:

\begin{numcases}
{P_n =}
|\mathcal{N}|^{2} N e^{-|\alpha|^2}\frac{|\alpha| ^{2n}}{n!} ,& for $n=SN$ \nonumber\\
\nonumber \\
0, & for $n \neq SN$
\label{PND final}
\end{numcases}

This is one of the main analytical results of the paper, where a close-form expression is obtained for the PND of a GCS. It is fascinating to observe that, the Poisson envelop is globally maintained, having contributions from the levels, $n=0,\; N,\; 2N,\; 3N,\; ...$.
We have also studied the variance of the state \emph{w.r.t.} the mean through the Fano factor \cite{dong2000superposition} analysis, where the Fano factor is defined as $\langle(\Delta n)^2 \rangle/\langle n\rangle$. Fano factor for higher $\alpha$ oscillates with $\alpha$ between sub- and super-Poissonian statistics. However, the sub-Poissonian nature becomes more probable for higher order mesoscopic superposition states.

\section*{Analytical Derivation of the PND for a GPACS}

We are going to focus on the properties of GPACS, which is produced by adding $r$ photons to each constituent state. Different protocols exist for realizing photon addition to a CS \cite{zavatta2004quantum,
li2018generation,shringarpure2019generating}. The general formalism is described by a GPACS, which takes the form
\begin{eqnarray}
|\psi,r\big\rangle =\frac{\mathcal{N}_{pa}}{\sqrt{N}} \sum_{j=1}^{N}|\alpha e^{\frac{i2\pi j}{N}},r\big\rangle = \frac{\mathcal{N}_{pa}}{\sqrt{N}}(a^\dagger)^r\sum_{j=1}^{N} |\alpha e^{\frac{i2\pi j}{N}}\big\rangle.
\label{PACSwavefunction}
\end{eqnarray}
The normalization constant $|\mathcal{N}_{pa}|$ is defined by
\begin{eqnarray*}
|\mathcal{N}_{pa}| = \left[\frac{r!}{N}\sum_{j,k=1}^{N}L_{r}
\left(-\alpha_{j}\alpha^{\ast}_{k}\right)exp(-|\alpha|^{2}
+\alpha_{j}\alpha^{\ast}_{k})
\right]^{\frac{-1}{2}},
\label{GPACSNorm}
\end{eqnarray*}
where $\alpha_{j}=|\alpha|e^{\frac{i2\pi j}{N}}$, $\alpha^{\ast}_{k}=|\alpha|e^{\frac{-i2\pi k}{N}}$ and $L_{r}
\left(-\alpha_{j}\alpha^{\ast}_{k}\right)$ being the Laguerre polynomial of order $r$.
We expand the coherent state into the Fock basis and operate the raising operator $(a^\dagger)^r$. The photon distribution is obtained by finding the contribution of the $n$-th level.
\begin{figure*}[htp]
    \centering
   \includegraphics[width=.99\textwidth]{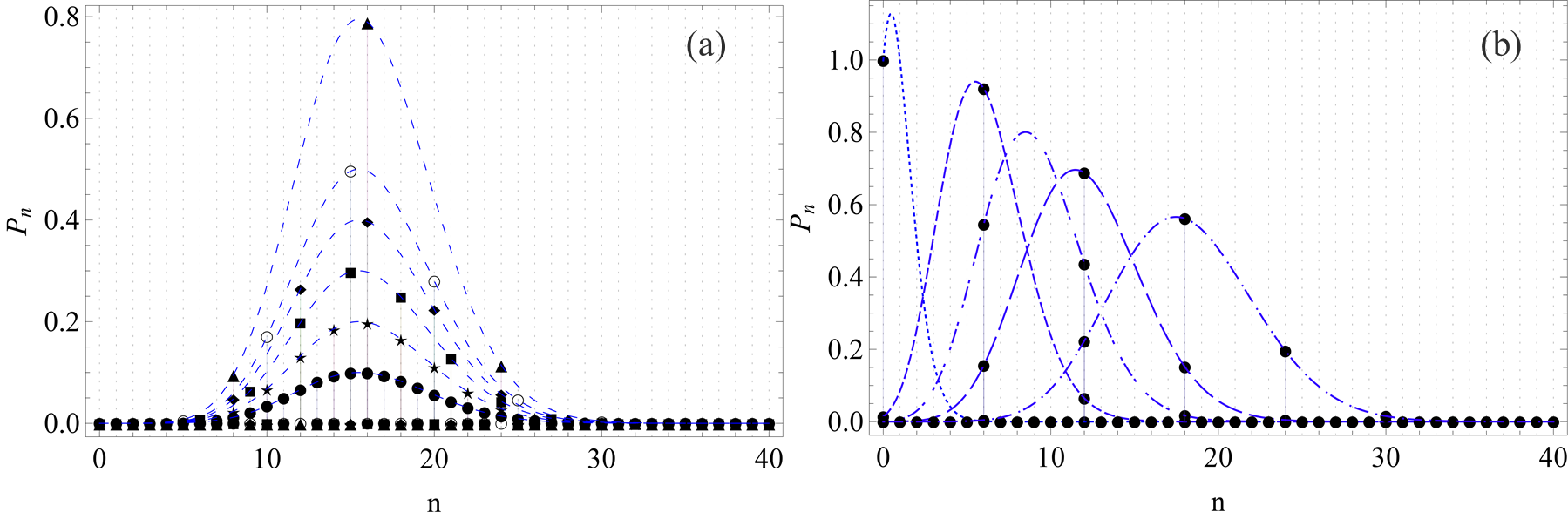}
    \caption{ a) PND of GCS with fixed $|\alpha|=4$ for a coherent state (dark circle), a Sch{\"o}dinger cat with $N=2$ (star), $N=3$ (square), Compass state with $N=4$ (diamond), $N=5$ (empty circle) and for $N=8$ (triangle). All the finite probabilities for getting photons lie on the Poissonian curves (dashed lines). b) PND of the quantum state with $N=6$ for varying $\alpha$: $|\alpha|=1$ (dot), $2.45$ (short dash), $3.0$ (dot-dash-dot), $3.46$ (larger dash), and $4.24$ (dot-dash).}
    \label{PNDany}
 \end{figure*}
We adopt similar derivation as the case of GCS and arrive at the final expression of PND for GPACS as

\begin{numcases} {P_n^r =}
|\mathcal{N}_{pa}|^{2} N e^{-|\alpha|^2}\frac{{|\alpha| ^{2(n-r)}}{n!}}{[(n-r)!]^{2}} ,& for $n=SN + r$ \nonumber\\
\nonumber \\
0, & for $n\neq SN + r$
\label{PNDPACSfinal}
\end{numcases}

Here, $S$ is a non-negative integer. The final expression of the PND is physically constituted by the contributions from $N+r,\; 2N+r\;...$ levels. Moreover, the PND of a GPACS gradually deviates from the Poisson envelop ($r=0$) with the addition of photons to the constituent states. The effect of photon addition on the variance of the state \emph{w.r.t.} the mean is also studied through the Fano factor. With increasing $\alpha$, it rises from zero and oscillates around unity. In this case, the sub-Poissonian nature becomes more pronounced for higher order mesoscopic superposition states and larger number of photon addition.

\begin{table*}[htpb]
\centering
\begin{tabular}{|c||c|c|c|}
\hline
    \multicolumn{1}{|c||}{Fock state } & \multicolumn{1}{c|}{GCS ($N$)} & \multicolumn{1}{c|}{GPACS ($N$)}  & \multicolumn{1}{c|}{Elliptical geometry}\\ 
       $|n\big\rangle$  & \;\;  &  \;\; &  (Both GCS and GPACS)\\
 \hline
             \;\;$|0\big\rangle$\;\; &  \;\;All $N$\;\; & \;\;All $N$ for $r=0$\;\; &    \;\;Null\;\; \\ \hline
             \;\;$|1\big\rangle$\;\; &      \;\;Null\;\;   & \;\;All $N$ for $r=1$\;\; &    \;\;Null\;\; \\ \hline
             \;\;$|2\big\rangle$\;\; &      \;\;Null\;\;   & \;\;All $N$ for $r=2$\;\; &    \;\;Null\;\; \\ \hline
    \;\;$|3\big\rangle$\;\; &      \;\;Null\;\;   & \;\;All $N$ for $r=3$\;\; &    \;\;Null\;\; \\ \hline
             \;\;$|4\big\rangle$\;\; &      \;\;Null\;\;   & \;\;All $N$ for $r=4$\;\; &    \;\;Null\;\; \\ \hline
             \;\;$|5\big\rangle$\;\; &      \;\;Null\;\;   & \;\;All $N$ for $r=5$\;\; &    \;\;Null\;\; \\ \hline
             \;\;$|6\big\rangle$\;\; &      \;\;Null\;\;   & \;\;All $N$ for $r=6$\;\; &    \;\;Null\;\; \\ \hline
             \;\;$|7\big\rangle$\;\; &      \;\;Null\;\;   & \;\;All $N$ for $r=7$\;\; &    \;\;Null\;\; \\ \hline
             \;\;$|8\big\rangle$\;\; &      \;\;Null\;\;   & \;\;All $N$ for $r=8$\;\; &    \;\;Null\;\; \\ \hline
            \;\;$|9\big\rangle$\;\; &      \;\;Null\;\;   & \;\;All $N$ for $r=9$\;\; &    \;\;Null\;\; \\ \hline
              \;$|10\big\rangle$\;  &      \;\;10\;\;     & \;\;$N=9$ to $6$;  $r=1$ to $4$  \;\; &    \;\;Null\;\; \\ \hline
              \;$|11\big\rangle$\;  &      \;\;11\;\;     & \;\; $N=10$ to $6$;  $r=1$ to $5$ \;\; &  \;\;Null\;\; \\ \hline
              \;$|12\big\rangle$\;  &      \;\;12\;\;     & \;\; $N=10$ to $6$;  $r=2$ to $6$ \;\; &    \;\;Null\;\; \\ \hline
              \;$|13\big\rangle$\;  &      \;\;13\;\;     & \;\; $N=12$ to $6$;  $r=1$ to $7$ \;\; &    \;\;Null\;\; \\ \hline
              \;$|14\big\rangle$\;  &      \;\;14\;\;     & \;\; $N=13$ to $6$;  $r=1$ to $8$ \;\; &    \;\;Null\;\; \\ \hline
              \;$|15\big\rangle$\;  &      \;\;15\;\;     & \;\; $N=14$ to $6$;  $r=1$ to $9$  \;\; &    \;\;Null\;\; \\ \hline
             \;$|16\big\rangle$\;  &      \;\;16\;\;     & \;\;  $N=15$ to $6$;  $r=1$ to $10$  \;\; &    \;\;Null\;\; \\ \hline
\end{tabular}
\caption{Production of the Fock states by superposing $N$-CSs and $N$-PACSs in circular and elliptic geometries from $|0\rangle$ to $|16\rangle$. A GCS can not produce a Fock state below $|n_c \rangle$, where $n_c=9$. For PACSs, $r$ signifies the number of photon addition. Elliptic geometry never generates a Fock state. The results are calculated with $99\%$ accuracy for taking $P_n\simeq 1$.}
\label{combinational table}
\end{table*}

\section*{Quantum Scissor Operations from the Obtained PNDs}

\subsection*{PND of a GCS and Methodology of Truncation}

It is worthy to start with the illustration of the derived modulated PND for one of the above cases. Particularly, the PND (normalized to unity) of the GCS from Eq.(\ref{PND final}) is depicted in Fig.\ref{PNDany}, where Fig.\ref{PNDany}(a) is shown for a fixed value of $|\alpha|=4$ and Fig.\ref{PNDany}(b) is shown for a given state, $N=6$. One can observe in Fig.\ref{PNDany}(a) that, the contribution from the energy levels in PND is governed by the integer multiple rule \textit{i.e}., $n=SN$ and the amount of contribution is guided by the Poissonian envelops, as obtained through the analytical derivation. A reduction of the number of contributing levels with increasing the number of constituent states, $N$ (from $N=1$ to $8$), amounts to the amplification of the Poissonian envelops, which helps to maintain the conserved photon number (decided by $|\alpha|^2$). Null contribution of the PND from $n\neq SN$ can also be understood directly from Eq.(\ref{PND final}). In the second case (Fig.\ref{PNDany}(b)), we demonstrate an $\alpha$-scanning of the PND for a benzene-like GCS ($N=6$) \cite{roy2009sub} for instance. For this particular state, one can involve higher energy levels by increasing  $|\alpha|$. The state will also comprise of more states for higher $|\alpha|$. For all such quantum states with different physical parameters, the PND will follow the analytically obtained Poisson envelop, thereby making it an universal entity. We have illustrated our result for a specific cases ($|\alpha|=4$ for Fig.\ref{PNDany}(a) and $N=6$ for Fig.\ref{PNDany}(b)), which is actually applicable for a wide quantum state combinations. In general, these states imply mesoscopic superposition states, which is widely used in quantum information processing. In Fig.\ref{numberbar}, we have demonstrated it for $N=16$. The specific $\alpha$-instances depicted through the vertical bars are relevant for the states, which will be explained in the following subsections. Notably, all the $\alpha$-values excepting such special instances produce mesoscopic superposition states of order-$16$, separability of which will increase with increasing $\alpha$. One of the salient spooky properties of quantum physics is nonlocal quantum superposition, which is the general representation of the states under consideration. Such mesoscopic superposition states are highly studied in various aspects of quantum technology and known to manifest sub-Planck phase-space interference structures, which provide the suitable tool for quantum metrology \cite{ghosh2006mesoscopic,ghosh2009sub,roy2009sub,
ghosh2014enhanced,ghosh2019sub,agarwal2022quantifying,
bera2020matter,bera2022quantum}. The required number of constituent mesoscopic quantum states is unravelled by the calculated PND. The analytical results for PNDs in Eq.(\ref{PND final}) and Eq.(\ref{PNDPACSfinal}), not only probe into the fundamental aspects of quantum optics, will also help to engineer various quantum states, which are relevant for quantum optics and quantum information.

\subsection*{Truncating to Fock States}

\begin{figure*}[htb]
    \centering
    \includegraphics[width=.85\textwidth]{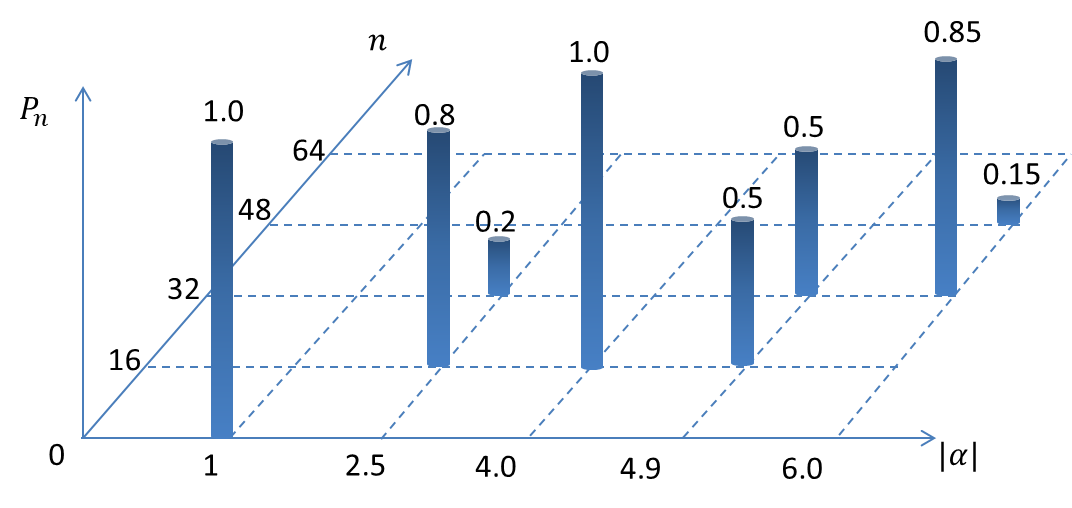}
    \caption{Important $\alpha$-instances of the PND for $N=16$ showing the generation of various quantum states. The numbers alongside the cylindrical bars are the corresponding values of the PND with particular $n$ and $|\alpha|$. $P_n$ is normalized to unity.}
    \label{numberbar}
 \end{figure*}

\begin{figure}[htp]
    \centering
    \includegraphics[width=.5\textwidth]{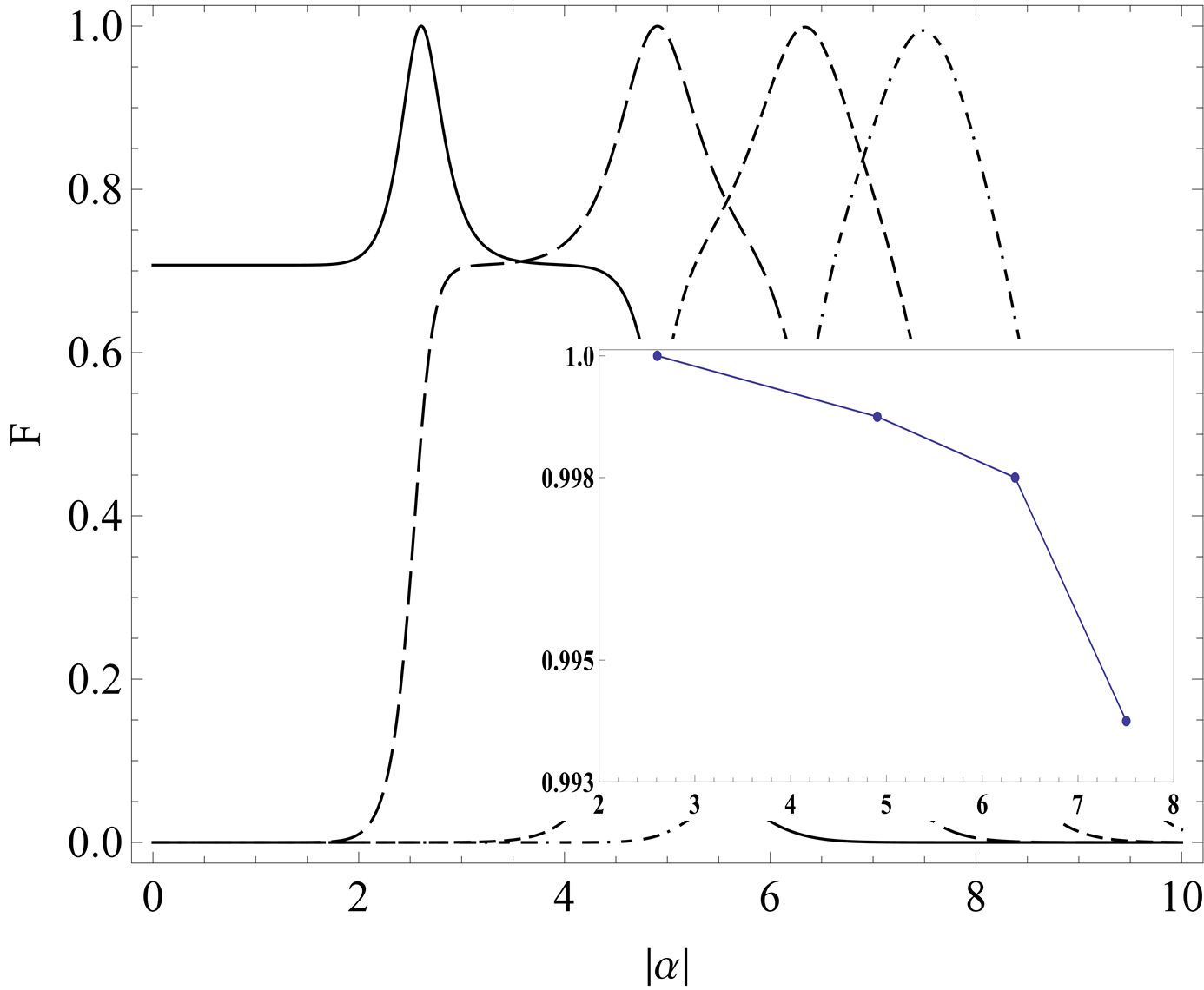}
    \caption{State overlap for GCS of order $N=16$ for equal superposed state with $S=0$ (solid line), $S=1$ (long-dash line), $S=2$ (short-dash line), and $S=3$ (dash-dot-dash line). Corresponding decrease of the overlap maxima for $S=0$, $1$, $2$ and $3$, is shown in the inset plot.}
    \label{Overlap}
 \end{figure}

In addition to the demonstration of the modulated PND for mesoscopic superposition states, we emphasize on the superposition of $N$ coherent states, which can produce Fock state. However, neither the said possibility is trivially understood, nor clearly reported till date. In this section, we will explore the same, which involves a crucial inter-relation between the number, type of the constituent states and the coherent state parameter. Here, we summarize the possibilities in tabular form (Table \ref{combinational table}) for both GCS and GPACS when the states are positioned along a circle or along an ellipse. Table \ref{combinational table} implies that a vacuum state ($|0\rangle$) can be produced from GCS of any order, but not GPACS. One can correlate the result from the bar-chart illustration in Fig.\ref{numberbar} for $N=16$, which shows the generation of Fock-states, $|0\rangle$ and $|16\rangle$, for which $P_n$ becomes unity with $|\alpha|=1$ and $4$, respectively. It is clear from column-II in Table \ref{combinational table} that Fock states $|1\rangle$-$|9\rangle$ can't be produced from a GCS, whatsoever be the experimental accuracy and tunability. On the contrary, these Fock states can be manufactured from GPACS (column-III) for $r=1$ to $9$, respectively. GCS again becomes capable of producing the Fock state beyond $n=9$. The latter domain is also relevant for a GPACS, where the fundamental condition becomes $N+r=n$. However, it is interesting to note from column-III that, producing Fock states, $|10\rangle$-$|16\rangle$, essentializes a lowest order of GPACS, which is $N=6$ or a benzine-like GPACS. Throughout the manuscript, the results are provided with $99\%$ accuracy of the analytically obtained PND. We have also examined the Fock-state generation with a deviation from the circular geometry and denoted in column-IV, particularly for an elliptic geometry, which conveys incapability of both GCS and GPACS to manufacture any of the Fock states.

\begin{figure*}[htpb]
\centering
      \includegraphics[width=.9\textwidth]{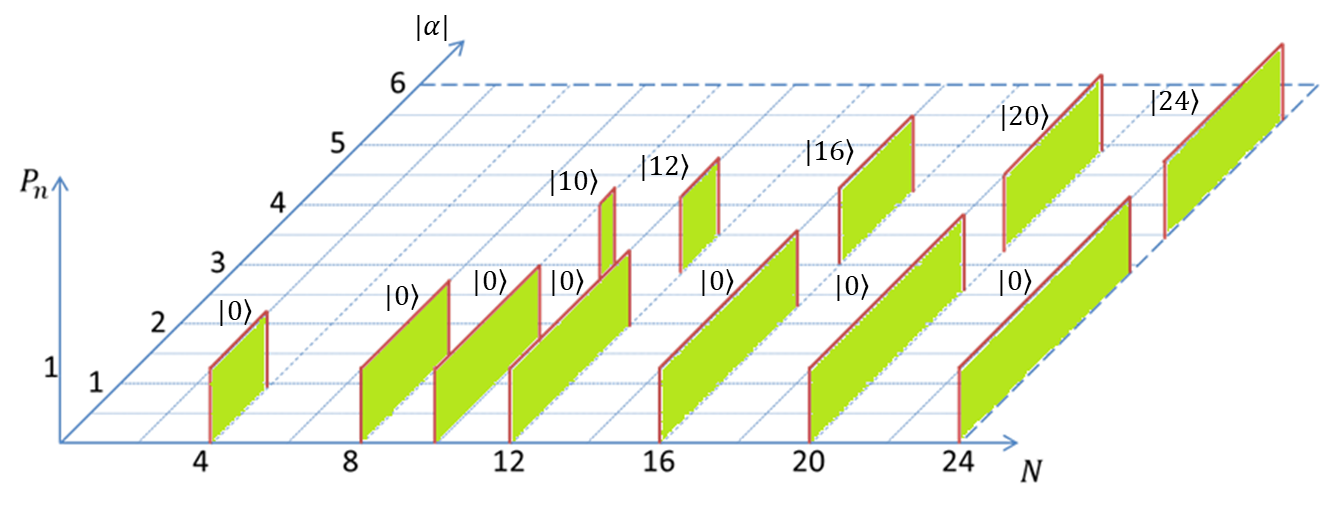}
\caption{Photon number window for generating Fock states, represented by the width of the rectangular sheets along $|\alpha|$-axis. $P_n =1$ situations are illustrated for discrete $N$'s: $N=4,\;8,\;10,\;12,\;16,\;20,\;24$, where $P_n$ is taken with $99\%$ accuracy. Vacuum state, $|0\rangle$, appears for all $N$ and the contribution from the next allowed level starts at $N=10$.}
\label{numberwindow}
\end{figure*}

\subsection*{Truncating to various Fock State superpositions}

Superposition of two Fock states is of paramount interest, especially due to their use as computational basis in quantum algorithms. As an illustration of a specific GCS for $N=16$ in Fig.\ref{numberbar}, we plot the analytically obtained PND upto $|\alpha|=6$. Linear superposition of two Fock states can be obtained for a variety of $\alpha$ in the form, $c_1 |SN\rangle +c_2 |(S+1)N\rangle$, such that $c^2_1+c^2_2=1$. We have depicted only three cases in Fig.\ref{numberbar} as follows.\\
$i)$ $|\alpha|=2.5$, for which the obtained state will be
\begin{eqnarray*}
|\Psi\rangle= \sqrt{\frac{4}{5}}\;|16\rangle + \sqrt{\frac{1}{5}}\;|32\rangle;
\end{eqnarray*}
$ii)$  $|\alpha|=4.9$, for which the obtained state will be a $50-50$ superposition:
\begin{eqnarray*}
|\Psi\rangle= 1/\sqrt{2}\;\;\left(|16\rangle + |32\rangle\right);\nonumber
\end{eqnarray*}
$iii)$ $|\alpha|=6$, for which the obtained state will be
\begin{eqnarray*}
|\Psi\rangle= \sqrt{\frac{17}{20}}\; |32\rangle + \sqrt{\frac{3}{20}}\; |48\rangle . \nonumber
\end{eqnarray*}
We have also studied the fidelity between the obtained state and the desired state for a equal Fock state superposition, such that the desired state is taken as
\begin{eqnarray}
|\psi^{\prime}\big\rangle = \frac{1}{\sqrt{2}}\big(|SN\big\rangle + |(S+1)N\big\rangle \big),
\label{superposed number}
\end{eqnarray}
whose scalar product with the GCS from Eq.(\ref{wavefunction}),
\begin{eqnarray}
|\psi\big\rangle=\frac{\mathcal{N}}{\sqrt{N}}  e^{\frac{-|\alpha|^2}{2}}\sum_{j=1}^{N}\sum_{n=0}^{\infty}\frac{\alpha^n}{\sqrt {n!}} e^{\frac{i2\pi jn}{N}}|n\big\rangle,
\label{wavefunction new}
\end{eqnarray}
becomes
\begin{eqnarray}
F  =&& \big\langle\psi^{\prime}|\psi\big\rangle \nonumber\\
   = && \frac{\mathcal{N}e^{\frac{-|\alpha|^2}{2}}}{\sqrt{2N}}\sum_{j=1}^{N}
  \Bigg[\frac{\alpha^n e^{\frac{i2\pi jn}{N}}}{\sqrt {n!}}\delta_{n,SN} +
  \frac{\alpha^n e^{\frac{i2\pi jn}{N}}}{\sqrt {n!}}\delta_{n,(S+1)N} \Bigg]\nonumber\\
   = && \frac{\mathcal{N}e^{\frac{-|\alpha|^2}{2}}}{\sqrt{2N}}\sum_{j=1}^{N}
  \Bigg[\frac{\alpha^{SN} e^{i2\pi jS}}{\sqrt {(SN)!}} +
  \frac{\alpha^{(S+1)N} e^{i2\pi j(S+1)}}{\sqrt {((S+1)N)!}} \Bigg]
\end{eqnarray}
We have provided the plots of this function in Fig.\ref{Overlap}, where $|F|^2$ is known as the fidelity. The maxima imply the values of $|\alpha|$, only for which the generations of equal superposition states may be possible. However, it is worth observing in Fig.\ref{Overlap} that, the maxima of the overlap $F$ diminish slightly for higher order GCS, as shown in the inset plot. For example, $1/\sqrt{2}\;\left(|0\rangle + |16\rangle\right)$ (solid line curve) has higher overlap maximum than $1/\sqrt{2}\;\left(|16\rangle + |32\rangle\right)$ (long-dash curve) and so on. Hence, the equal superposed state with highest fidelity becomes $1/\sqrt{2}\;\left(|0\rangle + |N\rangle\right)$ for any $N$.

\section*{Identifying the Parameter Domains for Realization}

In the previous sections, we have thoroughly discussed the possibilities for obtaining the quantum states in Table \ref{combinational table}, Fig.\ref{numberbar}, and Fig.\ref{Overlap} along with Eq.(\ref{PND final}) and Eq.(\ref{PNDPACSfinal}). However, it is extremely important to examine the allowed laser intensity parameter ($\alpha$) for their experimental generations. Below, we will elaborate the allowed domains of $\alpha$ for Fock states from a GCS, equal superposition of two Fock states from a GCS, and Fock states from a GPACS. Beyond these parameter domains, the quantum states (both GCS and GPACS) will display a mesoscopic quantum superpositions for larger $\alpha$ to make the constituent CSs distinguishable.

\subsection*{Parameter Domains for Truncating to Fock States from a GCS}

\begin{figure*}[htpb]
\centering
      \includegraphics[width=1.\textwidth]{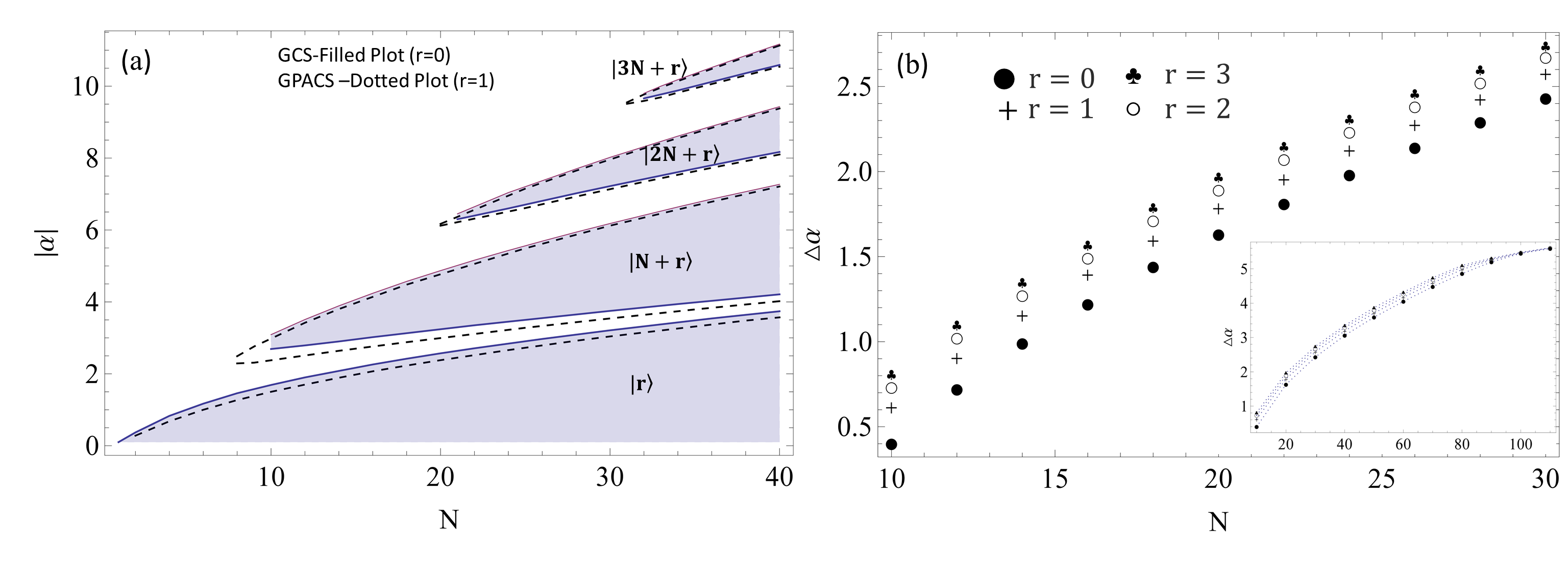}
\caption{(a) Graphical format of photon number window for the generation of Fock states $|r\big\rangle$, $|N+r\big\rangle$, $|2N+r\big\rangle$ and $|3N+r\big\rangle$. The shaded regions correspond to the GCS ($r=0$), whereas the region in between the pair of dotted lines correspond to GPACS ($r=1$). (b) Effect of photon addition ($r$) on the photon number window, $\Delta \alpha$, for $N$-th order superposition state, $|N+r \big\rangle$. $\Delta \alpha$ saturates for higher order superposition (inset plot) with varying $r$.}
\label{DeltaAlpha}
\end{figure*}

A numerical study of the PND for a large number of quantum states from Eq.(\ref{PND final}) is shorted out for $P_n=1$, to obtain the parameter domain for the Fock state $|n\rangle$. These photon number windows are represented in Fig. \ref{numberwindow} by the width (along $|\alpha|$-axis) of the rectangular vertical sheets for discrete $N$'s: $N=4,\;8,\;10,\;12,\;16,\;20,\;24$. Vacuum state, $|0\rangle$, appears for all $N$ with gradually increasing domain for higher order superpositions. However, generating the Fock state, $|N\rangle$, is possible only for $N\geq 10$, which is in agreement with Table \ref{combinational table}. A further analysis of the PND reveals that the Fock states, $|2N\rangle$ and $|3N\rangle$ \emph{etc}, are also possible for even higher order superpositions with $N\geq21$, $N\geq32$ \emph{etc}, respectively. This is better delineated in Fig.\ref{DeltaAlpha}(a) by the shaded regions with $r=0$. It is also interesting to observe that, the parameter domain for a specific Fock state gets wider with increasing $N$. However, for a given $N$, the window for $|2N\rangle$ is narrower as compared to the window for $|N\rangle$ and this trend continues. It is also intriguing to cite two specific cases from Fig.\ref{DeltaAlpha}(a): $N=12$ and $N=24$. First case implies that a superposition of $12$-CSs can produce vacuum state $|0\rangle$ for $0\leq |\alpha| \leq 1.9$ and also $|12\rangle$ for $2.79 \leq |\alpha| \leq 3.51$, but not the other higher order Fock state beyond $|12\rangle$. On the other hand, $24$-CSs can produce vacuum state $|0\rangle$ for $0\leq |\alpha| \leq 2.84$, $|24\rangle$ for $3.45 \leq |\alpha| \leq 5.43$, and $|48\rangle$ for $6.6 \leq |\alpha| \leq 7.04$, but not the other higher order Fock state beyond $|48\rangle$. Similar observation can be extracted from Fig.\ref{DeltaAlpha}(a) as and when required.

In the illustrations, we have taken the $P_n$ value with $99\%$ accuracy in the state overlap, regarding which, there are two aspects: i) the obtained analytical results are exact and require no accuracy. However, while trying to quantify the domain of $\triangle \alpha$ from the plot, we need $P_n=1$, but numerically it is never exactly one, as it includes machine precision. Hence, for demonstrating the results, one needs to uniformly fix an upper limit of $P_n$, beyond which we will conclude $P_n\sim1$, and consider obtaining the desired state. ii) Experimental accuracy is also an important factor. Depending on the situation, one can change this accuracy limit, which will make $\triangle \alpha$ to slightly vary in practical situation. The present model will be applicable to any such scenario, depending on the experimental accuracy, relying on several aspects, like detection efficiency, experimental environment, techniques \emph{etc}. Some related experiments have been performed to create $1$, $2$ or multiple photon Fock states with fidelity $70\%$-$95\%$ \cite{huisman2009instant,magana2019multiphoton,
cooper2013experimental}. Single photon source is also created with $99\%$ fidelity in a cavity QED \cite{zhang2018high}.

\subsection*{Parameter Domain for Truncating to Equal Fock State Superpositions}

We have already explained in the previous section that, the analytically derived PND is capable of identifying the $\alpha$-value also for an arbitrary linear superposition of two allowed Fock states. Equally superposed Fock state, having a special physical importance, can be generated for specific value of $|\alpha|$, allowed by Eq.(\ref{PND final}). The corresponding condition from Eq.(\ref{PND final}) becomes
\begin{eqnarray}
P_{n=SN}=P_{n=(S+1)N}
&&\Rightarrow  |\alpha| ^{2N}  =  \frac{((S+1)N)!}{(SN)!}\nonumber \\
\Rightarrow |\alpha| & = & \bigg({}^{SN+1}P{}_{N}\bigg)^{\frac{1}{2N}},
\label{pochhammer}
\end{eqnarray}
where ${}^{SN+1}P{}_{N}$ is the Pochhammer symbol, which is defined as ${}^{x}P{}_{n} = \frac{(x+n-1)!}{(x-1)!}$. One can solve $|\alpha|$ from Eq.(\ref{pochhammer}) for which a $50-50$ Fock-state superposition is obtained. This value depends on $ S $. For example, a GCS of order $N=16$ generates equal superposition of $|0\rangle$ and $|16\rangle$ for $S=0$ at $|\alpha|=2.6$; $|16\rangle$ and $|32\rangle$ for $S=1$ at $|\alpha|=4.9$ (shown in Fig.\ref{numberbar}); $|32\rangle$ and $|48\rangle$ for $S=2$ at $|\alpha|=6.34$; and $|48\rangle$ and $|64\rangle$ for $S=3$ at $|\alpha|=7.5$. We also verify that the maxima of the fidelity in Fig.\ref{Overlap} match with the obtained $|\alpha|$ values from this analysis, which is in conformity with the PND in Eq.(\ref{PND final}).

\subsection*{Parameter Domain for Truncating to Fock States from a GPACS}

Table \ref{combinational table} suggests that, unlike GCS, GPACS can produce Fock states, $|0\rangle$-$|9\rangle$, where the resultant Fock state is equal to the photon addition, $|r\rangle$. On the contrary, Fock states $|10\rangle$ onward are obtained when it is equal to $|N+r\rangle$. Hence, in the latter case, a particular Fock state can be produced with lesser number of PACSs, as compared to a GCS, which establishes an advantage to work with GPACS. The laser parameter domain is depicted in Fig.\ref{DeltaAlpha}(a) by the area between the pair of dashed curves for single photon added ($r=1$) GPACS. Regions for generating Fock states $|r\big\rangle$, $|N+r\big\rangle$, $|2N+r\big\rangle$ and $|3N+r\big\rangle$ are included till $N=40$. It is interesting to notice the deviation of the parameter domain with respect to GCS. The domains get broader with photon addition. The same is quite clear in Fig.\ref{DeltaAlpha}(b), where only $\Delta\alpha$ is plotted with $N$ for given photon addition $r$ (Here, $r=0,\;1,\;2,\;3$). $\Delta\alpha$ increases with the addition of photon, $r$, for a fixed $N$ and $\Delta\alpha$ also increases with $N$ for a fixed $r$. Inset plot in Fig.\ref{DeltaAlpha}(b) is delineated to manifest the saturation, occurred  for very large $N$. Merging the points of different $r$'s for very large $N$ signifies no effect of the photon addition on $\Delta\alpha$.

\section*{A Deviation from Circular Geometry: Elliptical State}
\begin{figure}[htb]
    \centering
    \includegraphics[width=.5 \textwidth]{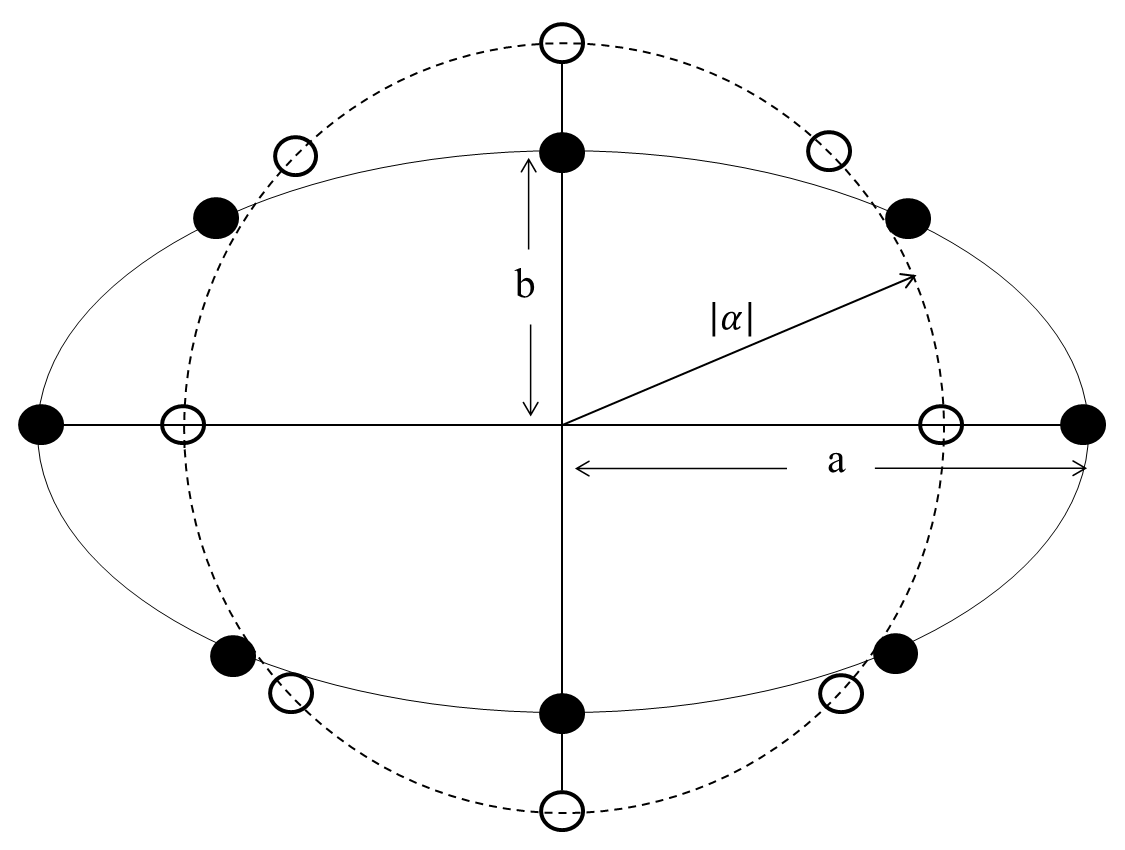}
    \caption{A deviation from the usual circular geometrical arrangement for a $N$-th order GCS. Circular and elliptical configurations are taken with equal area for comparison: $\pi |\alpha|^2=\pi ab$, with $a=5.0$, $b=3.2$ and $|\alpha|=4.0$.}
    \label{CoherentEC}
 \end{figure}

So far, we have been considering the mostly studied circular configuration for both GCS and GPACS. However, it is worthy to know whether there will be any significant change in physics by altering the geometry, as far as quantum state engineering is concerned. We have checked it for various geometries and found to have analogous results. Here, it is illustrated for an elliptic geometry. Elliptical states are studied in the context of quantum non-classicality \cite{wang2011nonclassical,miry2013nonlinear,
miry2014generation}. A quantum state of $N$-CSs, situated along an ellipse with equal phase difference, can be represented by
\begin{equation}
|\psi_e\big\rangle  = \frac{\mathcal{N}_{e}}{\sqrt{N}} \sum_{j=1}^N|\alpha_{j} e^{\frac{i2\pi j}{N}}\big\rangle
\label{Elliptiticalstate}
\end{equation}
where the distance of the $j$-th state from the origin becomes
\begin{eqnarray*}
|\alpha_{j}|=\left[\frac{\cos{\frac{2\pi j}{N}}}{a^{2}} + \frac{\sin{\frac{2\pi j}{N}}}{b^{2}}\right]^{-\frac{1}{2}}.
\end{eqnarray*}
$a$ and $b$ are real numbers, indicating the major and the minor axis of the ellipse, respectively. $\mathcal{N}_{e}$ is the normalization constant defined by
\begin{eqnarray*}
\mathcal{N}_{e} =\sqrt{N} \left[\sum_{j,k=1}^N e^{-\frac{(|\alpha_{j}|^{2}+|\alpha_{k}|^{2})}{2}} \exp\left({|\alpha_{j}|}|\alpha_{k}|e^{\frac{i 2 \pi (k-j)}{N}}\right)\right]^{-\frac{1}{2}}.
\label{EllipseNormalization}
\end{eqnarray*}

\begin{figure*}[htb]
    \centering
    \includegraphics[width=1.0\textwidth]{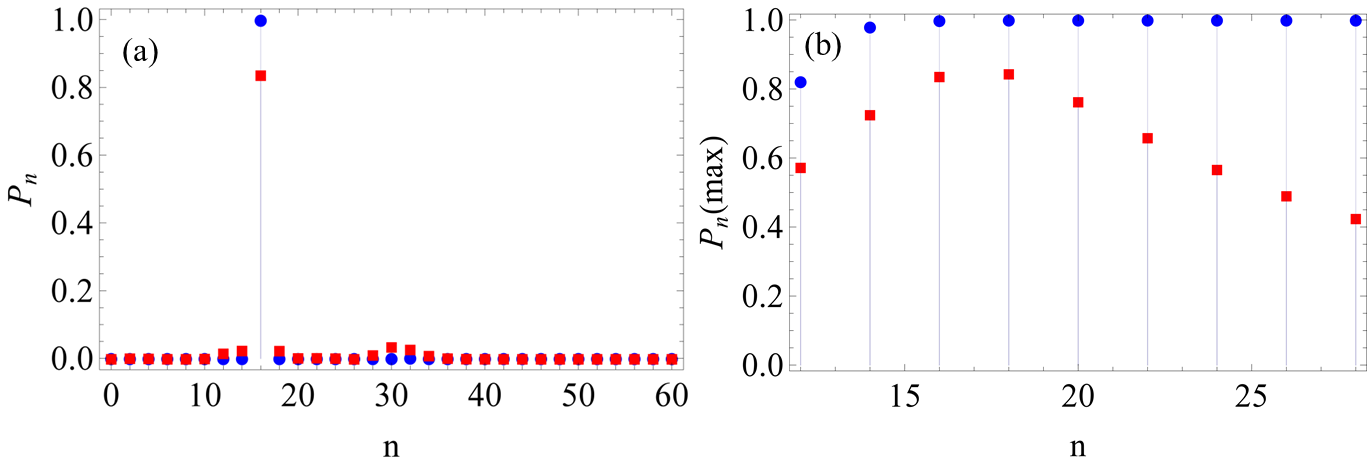}
    \caption{(a) PND for $N=16$, where $P_{n}=1$ for a circular GCS (filled circle) and $P_{n}=0.83$ for elliptical GCS (filled square). (b) Maximum value of the PND with respect to $N$, where the circular state $P_{n}$ saturates after getting maximum value at $n=16$ for $N=16$, while in elliptical state $P_{n}$ is reduced (filled squares) after $n=16$. Here, $|\alpha|$ is taken as $4.0$ with $a=5.0$ and $b=3.2$: $|\alpha| = \sqrt{ab}$.}
    \label{ECSEESCombined}
 \end{figure*}

Figure \ref{CoherentEC} shows the comparison between the circular and elliptical configurations by taking equal area: $\pi |\alpha|^2=\pi ab$, with $a=5.0$, $b=3.2$ and $|\alpha|=4.0$. To obtain the PND for the elliptical state (ES), we follow similar steps and write
\begin{eqnarray}
P_{n} && =  \frac{|\mathcal{N}_{e}|^{2}}{N} \sum_{j,k=1}^{N} e^{\frac{-(|\alpha_{j}|^{2}+|\alpha_{k}|^2)}{2}}  \left[2 \frac{|\alpha_{j}|^{n}|\alpha_{k}|^{n}}{n!} \right. \nonumber\\
&& \left. \times \cos\big(\frac{2 \pi n (j-k)}{N}\big)\right].
\label{Elliptitical}
\end{eqnarray}
The PNDs of the ES are plotted in Fig.\ref{ECSEESCombined}, corresponding to the configuration shown in Fig. \ref{CoherentEC}. We take the elliptic state, such that a circular state of CS-parameter $\alpha$ and elliptical state with major and minor axes, $a$ and $b$, have equal area: $(\pi |\alpha|^2=\pi ab)$. For $a=b$, the elliptical state changes to circular state. The condition ($|\alpha| = \sqrt{ab}$) will allow a variety of ellipse with many combinations of $a$-$b$ for a fixed $|\alpha|$. Here, we illustrate a particular case for $|\alpha| =4.0$ with $a=5.0$ and $b=3.2$. Figure \ref{ECSEESCombined}(a) shows the PND for $N=16$, where $P_{n}=1$ value appears at $n=16$ for a circular GCS (filled circle). However, the maximum value of PND for the elliptical GCS never reaches unity, but $P_{n}=0.83$, denoted by filled squares. An extension of this analysis is carried out for the maximum $P_n$ values with respect to $N$ in Fig. \ref{ECSEESCombined}(b), where the circular state $P_{n}$ saturates after getting maximum value at $n=16$ for $N=16$, while in elliptical state $P_{n}$-maximum (filled squares) goes down after $n=16$ for $|\alpha| =4.0$. Hence, we can infer that an elliptic configuration is incapable of producing any Fock states and their combinations, which ascertains the essence of circular configuration for quantum state engineering.

\section*{Conclusions}

We have reported a novel quantum scissor operation, mediated by the analytically derived photon distribution of both GCS and GPACS. We have demonstrated the scissor operations for generating vacuum State, Fock States, superposition of two Fock states and mesoscopic superposition states. The specific conditions and the corresponding allowed domains of laser intensity parameter are crucial for their experimental generations, which are addressed by the reported photon number statistics. We reveal that a GCS can produce a vacuum state, Fock states (beyond $|9\rangle$), and mesoscopic superposition states in different domains of $\alpha$. A GPACS supports all Fock state generations, unlike a GCS. The allowed laser intensity domain can also be controlled by photon addition. The underlying connection between a overcomplete basis (for CSs) and a complete basis (for Fock states) is physically  intriguing. We also establish the fact that circular arrangement will be beneficial in experiment. All the conditions and parameter domains, which are reported in this work, are fundamentally imposed by the quantum system. Hence, the present work provides a lot of insights on quantum scissor operations, paving diverse applications in quantum technology. This can be further applied to many useful cases beyond the demonstrated ones, such as to get three Fock-state superposition, other two-state superposition, and applications to quantum information and quantum computing.

\section*{Methods}

Analytical derivation of the photon number statistics for GCS and GPACS are performed. Then, these expressions are used for selecting contributing energy levels, profiled by the PNDs in a given parameter range and various quantum states for quantum scissor operations. The precise laser intensity windows for realizing a variety of quantum scissors are provided. In addition to demonstrating few cases, the importance of circular geometry is also delineated for efficient quantum scissor operations.

\section*{Acknowledgement}
The authors acknowledge Prof. G.S Agarwal for useful discussions. UR acknowledges the support from SERB-CRG, GoI. PKP and SG acknowledge support from DST (No. DST/ICPS/QuST/Theme-1/2019/2020-21/01), India.




\section{Data Availability}
All data generated or analysed during this study are included in this published article.

\section*{Author contributions statement}
All authors have equally contributed in the research and manuscript preparation.

\section*{Additional information}

\textbf{Competing financial interests:} The authors declare that they have no competing interests.

\end{document}